\documentclass[letterpaper,10pt,conference]{IEEEtran}
\IEEEoverridecommandlockouts

\usepackage{url}
\usepackage[hidelinks]{hyperref}
\hypersetup{breaklinks=true}
\usepackage{breakurl}

\usepackage{balance}       
\usepackage{graphicx}
\usepackage{amsmath}
\usepackage{subcaption}

\begin{document}

\title{\LARGE \bf
Curate, Connect, Inquire: A System for \textbf{F}indable \textbf{A}ccessible \textbf{I}nteroperable and \textbf{R}eusable (FAIR) Human-Robot Centered Datasets
}

\author{Xingru Zhou, Sadanand Modak, Yao-Cheng Chan, Zhiyun Deng, Luis Sentis, Maria Esteva\textsuperscript{*}
\thanks{This research was funded by Good Systems – Living and Working with Robots. 
The Texas Advanced Computing Center provided cloud computing resources.}%
\thanks{X. Zhou, S. Modak, Y. Chan, Z. Deng, L. Sentis, and M. Esteva are with The University of Texas at Austin, TX, USA. 
Emails: \{xingruzhou, sadanand.modak, yaocheng.chan, zdeng, lsentis\}@utexas.edu;}
\thanks{* Corresponding author: maria@tacc.utexas.edu}
}

\maketitle
\thispagestyle{empty}
\pagestyle{empty}

\begin{abstract}
The rapid growth of AI in robotics has amplified the need for high-quality, reusable datasets, particularly in human-robot interaction (HRI) and AI-embedded robotics. While more robotics datasets are being created, the landscape of open data in the field is uneven. This is due to a lack of curation standards and consistent publication practices, which makes it difficult to discover, access, and reuse robotics data. To address these challenges, this paper presents a curation and access system with two main contributions: (1) a structured methodology to curate, publish, and integrate FAIR (Findable, Accessible, Interoperable, Reusable) human-centered robotics datasets; and (2) a ChatGPT-powered conversational interface trained with the curated datasets metadata and documentation to enable exploration, comparison robotics datasets and data retrieval using natural language. Developed based on practical experience curating datasets from robotics labs within Texas Robotics at the University of Texas at Austin, the system demonstrates the value of standardized curation and persistent publication of robotics data. The system's evaluation suggests that access and understandability of human-robotics data are significantly improved. This work directly aligns with the goals of the HCRL @ ICRA 2025 workshop and represents a step towards more human-centered access to data for embodied AI. 

\end{abstract}

\section{INTRODUCTION}
The rise of AI-embedded robotics has made the need for high-quality datasets for varied training applications critical.  In response, researchers are increasingly creating datasets specifically for usage in AI applications. Derived from complex and often interdisciplinary studies using mixed research methods, these often large and multimodal datasets reflect both the robots' and the humans' perspectives; some gathered in the context of carefully designed experiments and others during observations in the physical world. However, despite the growing interest in creating and sharing data, the landscape of open human-robotics datasets remains uneven. 

To begin with, discovering these datasets is not straightforward. Many robotics datasets are hosted on platforms such as GitHub without permanent digital object identifiers (PDI), or in personal and laboratory servers ,occasionally behind restricted access mechanisms, without assurance of their long-term availability due to changes in servers and website maintenance. While more discover,able, many datasets published in institutional repositories with PDIs are not easy to reuse, as they are scantly described.  Because there are no agreed-upon standards about how much and how to describe the robots and their instrumentation, the participants, or the experimental conditions used to gather the data, the published datasets may not be understandable for other researchers to decide if and how to use them. In addition, HRI datasets involving human participants present varied ethical concerns. Since each published dataset has its own landing page, researchers have to examine them individually to determine if they are fit for reuse in their applications. The situation becomes more complex as more training data is needed, for which researchers have to review multiple datasets for possible integration. Not having the possibility to inquire and compare them at once is time-consuming. Either hosted on a lab server or on an institutional repository, storing, moving, and downloading large datasets is cumbersome and hinders their reuse.  

To address these challenges, we developed a system with three interrelated components: robotics data curation and publication in an institutional repository; a robotics knowledge graph to organize, relate, and integrate curated metadata; and a trained ChatGPT instance that allows context-aware access to multiple datasets via natural language interaction. The system is implemented across different reliable infrastructure components to assure the long-term sustainability and accessibility of the datasets. 

Based on experiences curating robotics datasets for different \href{https://robotics.utexas.edu/}{Texas Robotics} teams, we created a human-robot-specific data model to accurately represent the provenance, research methodology, and technologies involved in the development of HRI datasets. The data model is implemented as a knowledge graph running on the Texas Advanced Computing Center’s (TACC)\cite{TACC2024} cloud infrastructure. In tandem, we developed a data report template that researchers can use to document the data model elements .  Datasets are curated and described by their creators according to the guidance offered in the template, and they are uploaded to the Texas Robotics Dataverse \cite{RoboticsDataverse2024} at the Texas Data Repository (TDR)\cite{TDR2024}.  Once datasets are published, their metadata is automatically harvested from the repository, mapped to the data model elements, and integrated into the corresponding nodes in the knowledge graph, enabling a normalized description across different datasets and thus their comparison. The knowledge graph schema and metadata, the data reports, and the datasets-related publications are used to feed into a ChatGPT-based chatbot, allowing users to query and retrieve data using natural language through a conversational interface. Currently in prototype mode, the system has seven registered datasets generated through different robotics studies and published in the Texas Robotics Dataverse\cite{Chan2024a, Chan2024b, Zhang2023, Gupta2024, Norman2024, Sharma2024, Karnan2022, Knox2023}. Figure 1 shows the system's components and workflow. 

To assess the system, we carried out different evaluations. We designed an expert assessment around information targets to identify if the chatbot's answers are consistently reliable, and we conducted a think-aloud session with a robotics expert naive to the registered datasets to observe how he interacted with the system and his satisfaction with the outcome.  Finally, we checked the entire system against the FAIR curation principles to assess whether data are Findable, Accessible, Interoperable, and Reusable (FAIR)\cite{Wilkinson2016}. The results suggest that the system retrieves accurate information, that it aids data discovery and exploration, and that it facilitates comparison between datasets. Our contribution highlights the importance of data curation and structuring to train a reliable LLM (Large Language Model). It also emphasizes the importance of a solid infrastructure to address reliable inquiry and continuous access to robotics datasets.  This work has the potential to promote robust data curation practices within our research community. The system represents a step towards a FAIR human-robotics data ecosystem. This work is aligned with the goals of the HCRL @ ICRA 2025 workshop. In particular, by addressing the challenges of data accessibility.  
\begin{figure}[t]
  \centering
  \includegraphics[width=1.0\columnwidth]{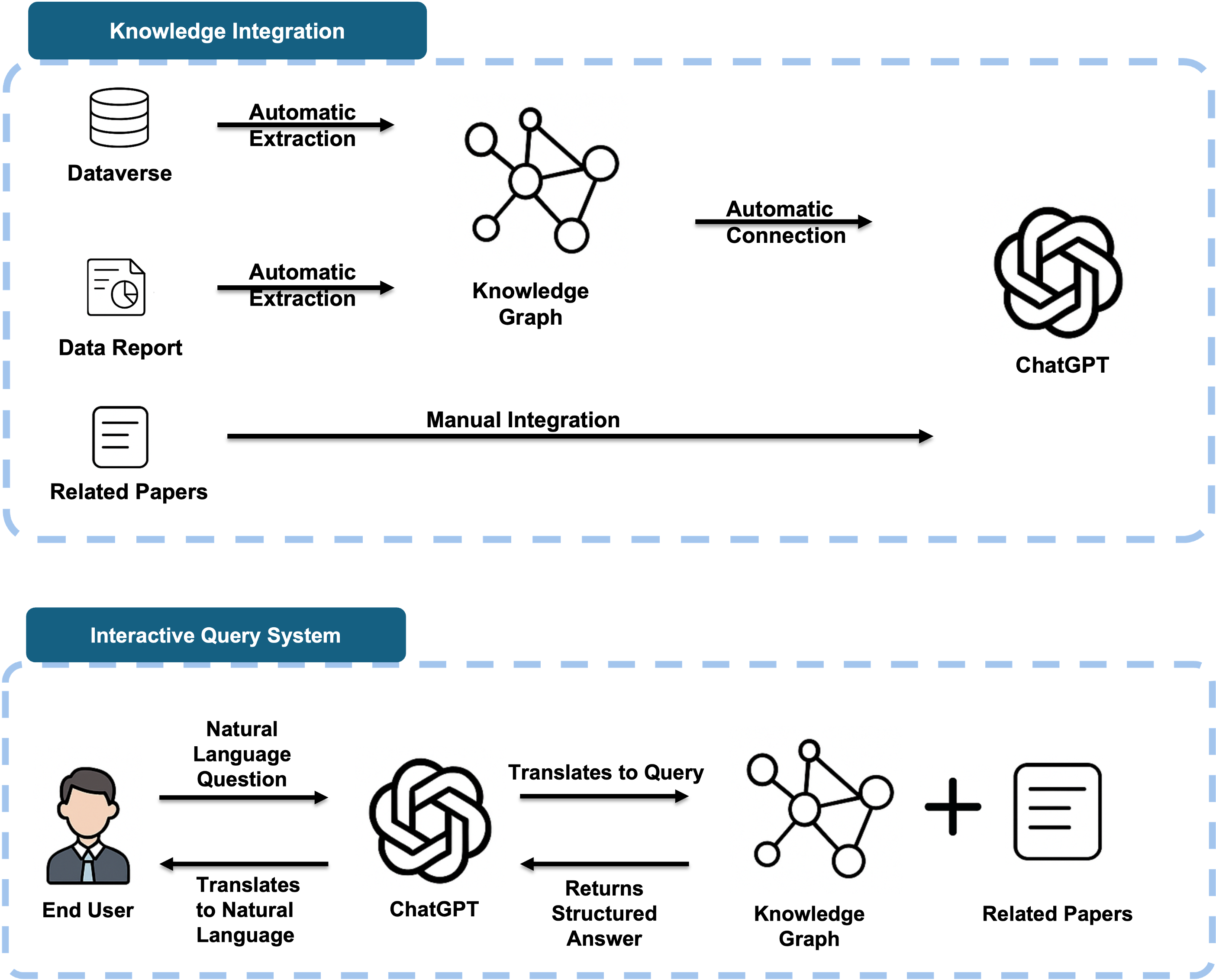}
  \caption{Schematic of the system's curation and access workflow}
  \label{fig:automated_knowledge_retrieval}
\end{figure}
    
\section{The Landscape of Large Human-Centered Robotics Datasets}

While most recently, researchers are producing a variety of robotics datasets, data curation and publication are still emerging practices in the robotics community. In the \href{https://www.re3data.org}{Registry of Research Data Repositories (re3data.org)}, which maintains a list of data repositories worldwide, there are no domain-specific repositories for robotics datasets, and currently there are no shared metadata schema and best practices to curate and publish them.  As a result, robotics datasets are scattered across different platforms, inconsistently described and often hard to understand and access. Researchers looking for reusable robotics data often need to search multiple platforms, including GitHub, Zenodo\cite{Zenodo2024}, or personal lab websites, which makes the process slow and unreliable. Even in institutional repositories, search results are often unsatisfactory due to poor metadata or missing documentation. For example, a search in the general-purpose repository  Zenodo using the term ``Human Robot Interaction", retrieves 70 datasets which have to be examined one by one to know their purpose and characteristics. While some datasets, such as AFFECT-HRI\cite{Heinisch2024} and HRI-CUES\cite{Irfan2024}, are fairly well documented, the majority lack basic information, such as how the data was collected. 

P2PSTORY \cite{P2PSTORY2025} from MIT Media Lab, UE-HRI \cite{UEHRI2025}, and PInSoRo \cite{PINSORO2025} are datasets stored within university websites and on GitHub. Because these platforms are not data repositories, the datasets lack PIDs, and there is no guarantee of their long-term sustainability. For example, the dataset associated with the Deep Fingerprinting project \cite{Sirinam2018}, initially hosted on \href{https://github.com/deep-fingerprinting/df}{GitHub project}, is no longer accessible through the provided download link.\footnote{See GitHub issue: https://github.com/deep-fingerprinting/df/issues/35} This illustrates how researchers and students may move on, and websites change. Lacking the infrastructure needed for permanent preservation, many datasets are at risk of becoming inaccessible. 

Another accessibility roadblock is the size of modern robotics datasets containing large numbers of heavy Rosbags and other complex image files. These are difficult to manage and access via a web browser. Most repositories, such as Zenodo\cite{Zenodo2024}, accept datasets of up to 50 GB to 1 TB, and GitHub will only hold up to 100 MB\cite{GitHubLargeFiles2024} per project. Across the board, what is missing are shared best practices for curating HRI datasets in ways that support long-term, cross-domain, and ethical use. Without this, valuable datasets are at risk of being lost or underused. Our work is motivated by this gap and seeks to offer a practical solution for improving how robotics datasets are optimized for reuse in the context of large-scale, human-centered learning.

\section{Components of the FAIR Data Curation and Access System}

 Modern HRI experiments and real-world robot observations entail complex study designs and cutting-edge technologies. Consequently, the derived datasets are multimodal and structurally intricate, and the involvement of human subjects in the studies adds another layer to ensure ethical data publication. Therefore, curation of HRI and AI-embedded robotics datasets demands a thoughtful, reproducible approach that captures the complexity of interactions transparently and ethically. To support this, we developed a system encompassing curation and access whose components we describe in the next sections using as case studies human-centered datasets published in the Texas Robotics Dataverse. To guide the direction and components of the system, we use the FAIR principles, a set of standards that address requirements for curation and publication of datasets and for the infrastructure that hosts them.

\subsection{A Uniform Data Model for Robotics}

Through the process of helping Texas Robotics researchers to organize their data, and hearing about their studies and how they collect and process data, we created a hierarchical data model as an abstract representation of human-centered robotics datasets. The model defines a set of core classes and properties, as metadata elements - that reflect common components of different studies from which robotics datasets derive. Representative classes and properties include, for example. \textit{robot type - robot model -robot equipment/sensor- robot control; research method- experiment location - experiment settings - experiment session - experiment condition}, etc. Because all curated datasets conform to this shared model, the resultant metadata for each dataset will be internally consistent and generalizable across all, making them interoperable. This interoperability enables scalable integration and comparison of datasets from different sources. 

\subsection{Curation Challenges and Recommendations, Metadata Standards, and the Data Report Template}

Data curation is at the system's foundation. Curation encompasses best practices for data organization and description, ethical publication, and infrastructure to ensure long-term sustainability and accessibility. \cite{DCC2025}. Since there are no specific metadata standards or curation guidelines for robotics data, we gained experience by following general curation best practices, by observing how existing datasets were publicly released\cite{OpenX2023, DROID2023, VertiWheelers2020, RH20T2023, DexYCB2021}, and through the process of curating and publishing different types of datasets for \href{https://robotics.utexas.edu/}{the Texas Robotics} research groups.

Data is deposited in the TDR, a general-purpose institutional repository at the University of Texas Libraries that provides long-term preservation, persistent identification through DOIs, and public access to datasets created by researchers from a consortium of universities in the state of Texas. To avoid their dispersion among datasets from different disciplines, we curate and publish the robotics datasets within a Texas Robotics Dataverse.  The baseline metadata for describing and representing the datasets in the repository is provided by the \href{https://dataverse.org/ }{Dataverse Project}, which is the underlying open source repository software for the TDR.  Among other metadata standards, Dataverse adopts the Data Documentation Initiative (DDI) schema\cite{DDI2021}, designed to describe Social Science datasets. DDI offers the possibility to include a high-level description as well as specific social science information, which is useful to describe the human subjects component of an HRI dataset. DDI does not, however, have elements to describe robotics-specific technical provenance needed for researchers to decide whether they can reuse it.  As researchers deposit data they fill in the DDI metadata fields. This metadata is  formatted as a JSON file that can be downloaded from the repository once a dataset is published. Using an open source repository assures that data is findable, as the standardized metadata is exposed to search engines and academic aggregators via standard protocols. It also assures data interoperability as the standardized metadata can be exchanged across repositories.  

To capture more in-depth robotics information about the datasets and to guide researchers in their curation process, we designed a data report template. The template is related to the elements in the robotics data model. Therefore, from noting the robot's model and its sensor equipment, to describing the experimental or observational methodology and the participants' tasks and behavioral or physical measures, to explaining the data post-processing methods (e.g. segmentation and labeling), all critical aspects about the datasets are included in the template as descriptive elements. This information ensures understandability and transparency, and structured semantic integration in the downstream knowledge graph. 

As we curate new datasets and encounter new elements that need to be described, we include them in a dedicated data report appendix developed to track emerging patterns. As specific elements appear more regularly, we promote them into the body of the data report and into the data model. This iterative strategy allows the system to grow and adapt while moving towards broader standardization. The approach supports research reproducibility as well as interoperability between datasets. The data report is included in the dataset publication in PDF format and used both for metadata extraction into the knowledge graph and as a source document in the Retrieval-Augmented Generation (RAG) pipeline to support accurate, context-aware responses from ChatGPT.

\subsection{Data Quality}
Unlike journals or conference proceedings, institutional repositories are self-publishing entities and do not have peer review in place. Thus, it is up to the researchers and curators to demonstrate a dataset's quality. Included in the data report template is a data quality statement section to record the types of quality control activities performed prior to releasing the datasets. Quality control items include standardized data collection (with consistent conditions and sensor calibration), annotation accuracy (verified through multi-step review and inter-rater reliability checks), and data integrity (ensured through automated and manual validation). In the case of datasets created to train models, we request that the location of the models/software is referenced, preferably with DOIs, and that the results of the datasets' performance become part of the documentation \cite{Zhang2023, Knox2023}. Data report guidelines addressing data quality also include using open source file formats for long-term preservation and requesting the inclusion of data dictionaries to explain variables in tabular data.  Comments from users in relation to the robotics datasets publication's quality and completeness are received via the feedback form in the datasets' landing pages, and we incorporate those as elements in the template. 

\subsection{Dataset Ethics}
Ethical aspects involving human subjects are carefully gauged and discussed with researchers at the point of study design and included in the report.  Considering compliance with IRB decisions for data anonymization and access restrictions, different strategies can be adopted. In the case of  CODa, recordings of incidental participants were removed upon request, and in the Community Embedded Robotics dataset\cite{Ryan2025},  participant faces were not included in the published video data. Because in many cases facial expressions are important to capture for research purposes, in \cite{Chan2024b} researchers sought informed consent, and all but one participant were comfortable with having their session recording released to the public without face blurring. 

Interdisciplinary teams may exhibit different opinions about privacy and data sharing. During the Robot Encounter \cite{Ryan2024} study, in which participants wearing physiological sensors to measure stress levels shared a common space with robots, social scientists had concerns about sharing the full text of focus groups, fearing that the participants' identity may be recognized. Instead, roboticists considered that anonymized  ECG and EDA recordings could be openly shared pending the participants' consent. The resultant dataset publication includes open sensor data but only excerpts and themes resultant from the focus groups. Acknowledging the need to find a common ground for sharing human subjects’ data, we identified topics that need to be reckoned with by interdisciplinary teams at the design phase of an HRI study.  These include a) analyzing the degree of disclosure and sensitivity of the interview topics and potential responses, b) considering the privacy risks of all the data types that will be recorded about participants,  and c) requesting participants’ consent for sharing each type of data. In the data report, we also require that all human subject research instruments, including surveys, questionnaires, interview protocols, and code books, be published to provide adequate context.

\subsection{Scalable Organization and Access for Large Robotics Datasets}

\begin{figure}[t]
  \centering
  \includegraphics[height=0.3\textheight]{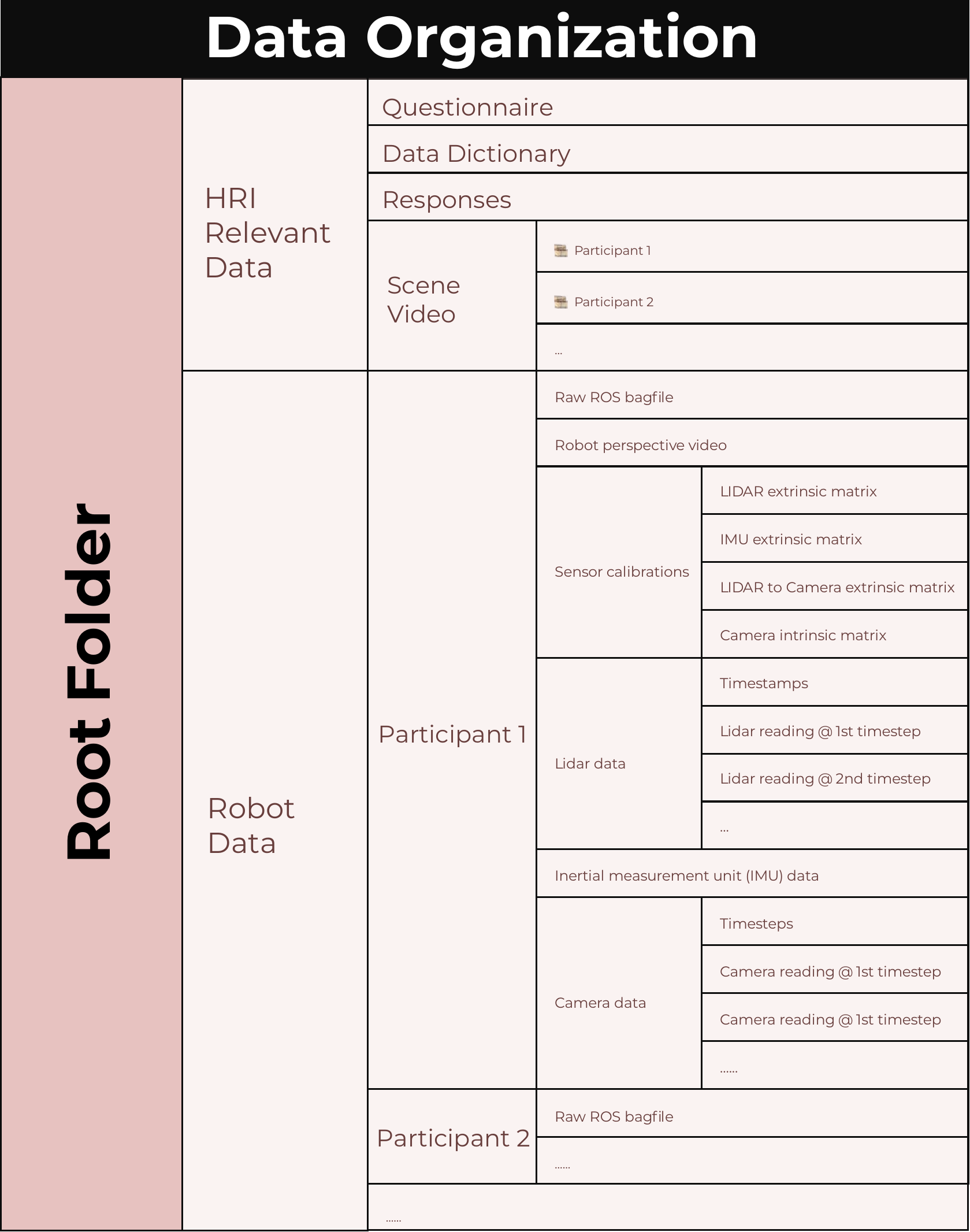}
  \caption{Vid2real Real\_World Collection Structure and Naming Convention}
  \label{fig:structure}
\end{figure}

The size of a dataset is relevant to its understandability and accessibility regarding how data is organized and whether it can be downloaded with ease. In terms of data organization, we provide guidance on folder and files organization and naming conventions that reveal the content of the files and are in alignment with the robotics data model elements. This alignment supports consistent labeling of experiment sessions, participant roles, and robot modalities, ensuring both human and machine-readable consistency for downstream indexing and retrieval. This is especially important to help users navigate large multimodal datasets derived from experiments with multiple testing sessions, or involving repeated observations with multiple recording instruments.  The Dataverse software allows tree views of the hierarchy that reflect the dataset's organization as well as the possibility to add descriptions to all data files, improving understanding and accessibility of the dataset. The data organization and file naming convention schema have to be described in the data report. An example of a documented file naming schema is shown in Figure \ref{fig:structure}. The schema is critical for machine processing, as this defined organization directly informs the mapping of the dataset's components to the data model within the knowledge graph and facilitates automated metadata extraction from the data report. The explicit organizational schema provides context from the knowledge graph to the LLM, facilitating accurate natural language retrieval of specific, knowledge graph-linked files. 

Training datasets are often bigger than the 1TB size limit allowed by the repository. To comply with scalable storage, long-term preservation, and ease of access, we integrated the Texas Robotics Dataverse with a high-performance web-accessible storage resource deployed at the Texas Advanced Computing Center at \href{https://tacc.utexas.edu/systems/corral/}{TACC} to host large-scale collections. This approach is used to store the 4 TB CODa dataset\cite{Zhang2023}.  This hybrid approach facilitates finding the datasets online through the repository's search engine optimization strategy while enabling permanent storage, scalability, and accessibility.  Part of the curation process entails developing scripts for automated download of large datasets both from TDR and from TACC's storage resource. Prepared by researchers in relation to their data organization, the scripts allow downloading all or particular portions of large datasets.

\subsection{Semantic Integration through a Knowledge Graph}

Once the data is published, the metadata is mapped to classes and properties in the robotics data model  implemented in the Neo4J-based knowledge graph.  The graph converts individual metadata records into interconnected networks of nodes and relationships facilitating advanced reasoning, filtering, and the effective preparation and contextualization of data for training machine learning models. The modeling approach enables queries that go beyond basic search functions. For instance, since robot models are structured as nodes in the knowledge graph, users can ask, “Which datasets use Boston Dynamics Spot?” and retrieve specific answers. This graph structure reflects metadata best practices seen in other domains like biology or geoscience\cite{Reese2022}, \cite{Hoyt2022}, \cite{Janowicz2024}, where standard schemas allow complex relationships and rich semantic queries. By requiring researchers to report consistently across robot types, experimental design, and human subject details, the system builds a trustworthy base for LLM interaction (key to avoiding hallucination or degradation in responses due to missing or inconsistent data) and facilitates data reuse.

\begin{figure}[t]
  \centering
  \includegraphics[width=1.0\columnwidth]{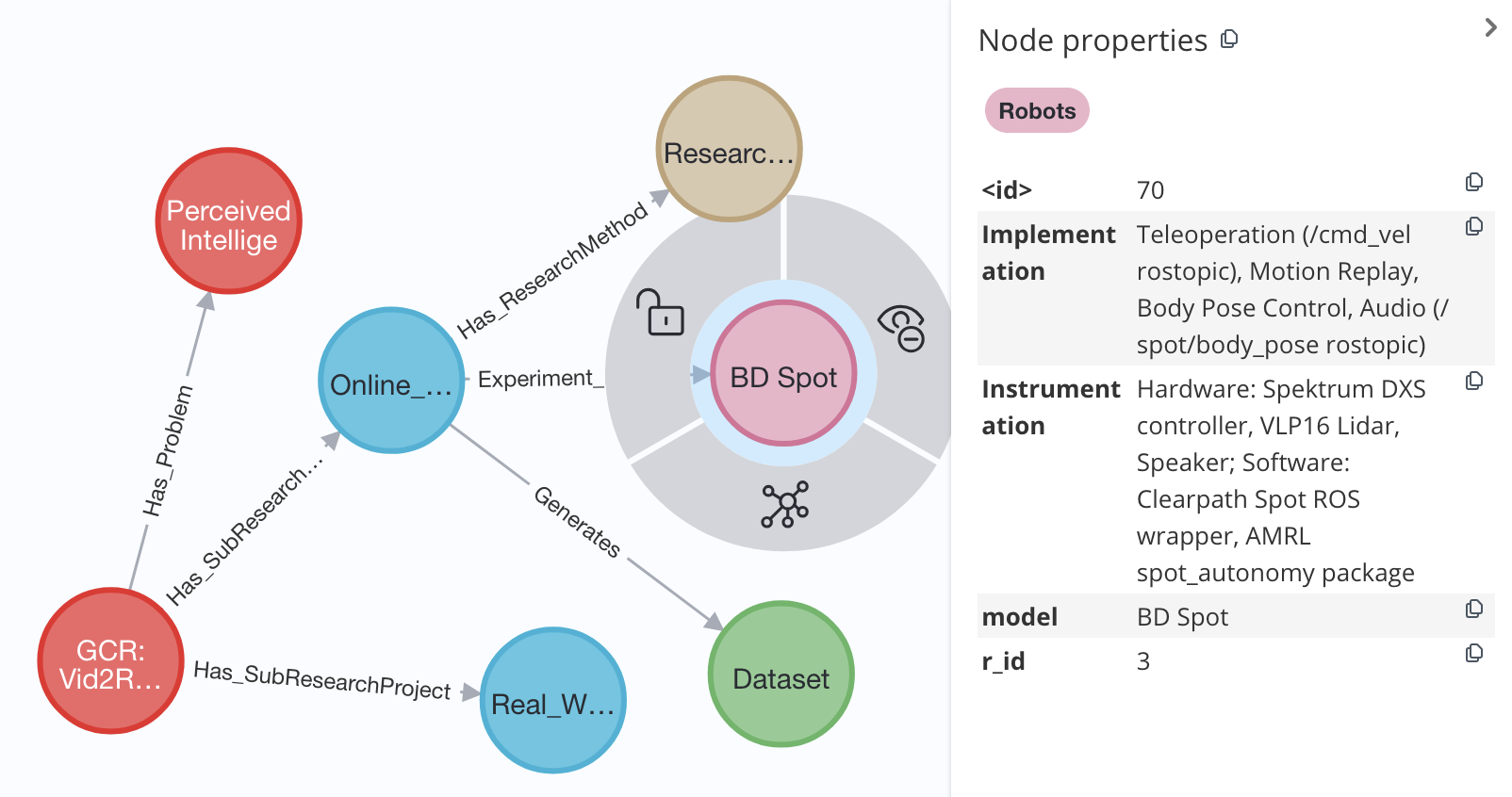}
  \caption{Vid2real Online Study Robot Metadata Class/Node and corresponding properties in the knowledge graph.}
  \label{fig:neo4j_structure2}
\end{figure}

Metadata extraction and knowledge graph population currently rely on the structured JSON metadata records from the Texas Robotics Dataverse datasets and the information input by researchers in the data report template.  Python scripts were designed to parse the structured fields within these data reports—fields intentionally aligned with our robotics data model— and to process the DDI-based JSON records. The scripts utilize pattern matching and keyword detection (such as identifying terms like ``robot'', ``participant'',  ``robot model'', "experiment session", "interview", "survey", "condition", etc.) to locate relevant metadata elements. These elements are systematically mapped to corresponding node types within Neo4j, ensuring precise and consistent semantic structuring. For example, a metadata field such as “Robot Model: Boston Dynamics Spot” becomes a node labeled RobotModel, linked to its parent dataset node through a defined \texttt{usesModel} relationship, which signifies the specific robot model utilized in the study that generated the dataset. This allows higher-level semantic inference and structured querying across datasets—capabilities that are not possible with flat or unstructured metadata. Figure \ref {fig:neo4j_structure2} demonstrates this mapping approach using as an example from \cite{Chan2024a}. 

\subsection{Human-Centered Access via an LLM}

The final layer of the system connects the knowledge graph to an interactive chatbot powered by an LLM using RAG. Instead of relying solely on pre-trained knowledge, the chatbot retrieves structured metadata from the Neo4j knowledge graph, combined with other relevant materials such as related publications and data collection instruments.  This comprehensive context is intended to improve the chatbot's delivery. 

Researchers can query the datasets through natural language, asking questions about one or more datasets, such as ``What robot model was used in the Vid2Real online study?'', ``Which studies use a Boston Dynamics robot?'' ``Does the CODa dataset include LiDAR ?'', ``How does the online study compare to the real-world study in terms of participant experience?'' ``Which research methods are used in the online and real-world Vid2Real studies?''. Using the RAG framework the ChatGPT is paired with structured responses drawn directly from the trained data. As a result, the way in which users can interact with this system is very different from the typical keyword search in Google, GitHub, or Zenodo. The richness and the structure of the curated metadata enhance factual grounding and mitigate the risk of hallucinated or overly generalized answers. It also allows retrieving specific files by asking questions such as ``Point to all video files for session 1 in the Vid2Real real world study''.

\section{Evaluation: Assessing the Performance of the Chatbot System}

We conducted two evaluations : (1) an expert review of the chatbot's performance, and (2) a pilot think-aloud session with a roboticist to assess the system's practical utility.

\subsection{Expert Review of Chatbot Information Quality}

The evaluation was conducted using the following curated datasets \cite{Chan2024a,Chan2024b}, \cite{Zhang2023}, \cite{Gupta2024,Norman2024}, \cite{Sharma2024},  \cite{Karnan2022}. These datasets originated from five different robotics laboratories and cover distinct HRI objectives—multimodal perception, social navigation, online video annotation, and object‑centric mapping—giving the system's prototype a heterogeneous and realistic testbed. Experts rated the chatbot’s performance against four dimensions \cite{Wu2024, DiBartolomeo2023}. Each dimension was tested using a set of ten task-specific natural language questions.  The chatbot/s responses were assessed on four dimensions: accuracy, consistency, comparison capability, and information completeness, each of which was rated on a 0–5 scale. The experts were the creators or curators of the datasets, which equips them with deep, first-hand knowledge necessary to judge the veracity and completeness of the chatbot's responses.
The four dimensions are: 
\textit{Information Retrieval}: Ability to provide complete, relevant, and clearly structured answers to user queries regarding specific dataset content. Example queries included, ``Provide me with the questionnaire link for the VID2REAL real-world study'' and ``List the robot behavioral conditions in the VID2REAL study.''

\textit{Answer Stability}: Consistency of responses across semantically similar queries phrased differently. Examples of tested queries include, ``What kind of robot is used in Vid2Real Real World?'' versus ``What type of robot was utilized in the Vid2Real real-world study?'' and queries posed in multiple languages like Chinese.

\textit{Factual Accuracy}: Correctness and precision of the information provided based on the underlying metadata. Queries assessed details such as IRB approval, sensor types, robot models, and specific experimental setups. 

\begin{figure}[t]
  \centering
  \includegraphics[width=1.0\columnwidth]{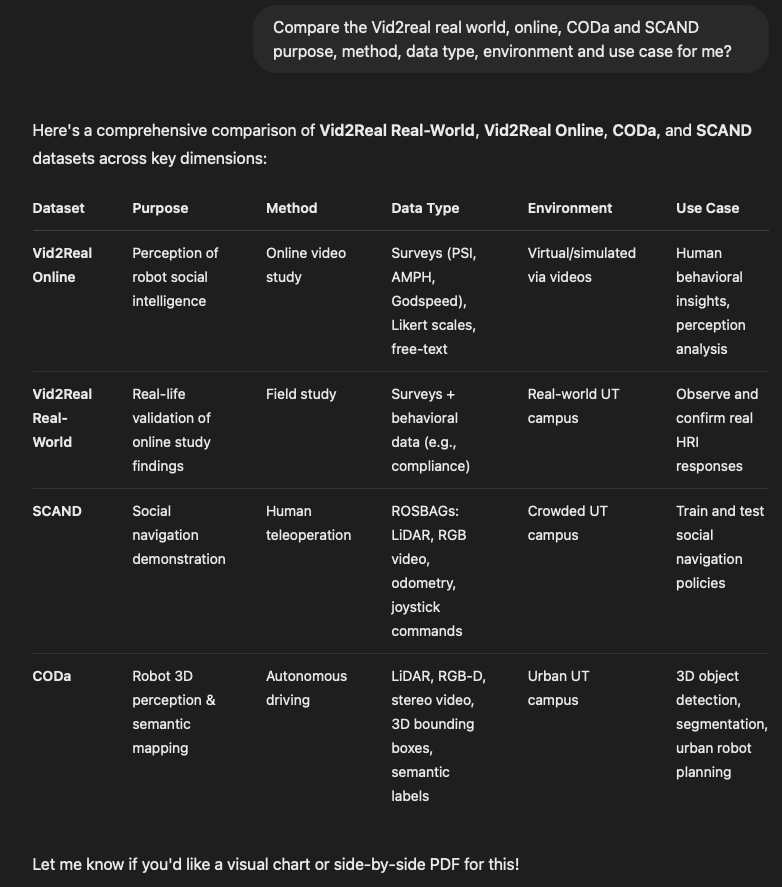}
  \caption{Example of Evaluation: Comparison Capability}
  \label{fig:compare}
\end{figure}

\textit{Comparison Capability}: Ability to accurately compare two or more datasets, highlighting their key differences. Example queries included, ``What are the differences between Vid2Real Real World and Vid2Real Online studies regarding robot control and sensory input?'' and ``Which datasets use joystick-based teleoperation versus autonomous navigation?'' An example comparison query and the chatbot's answer are shown in Figure \ref{fig:compare}.  

To reduce individual rating biases, we applied a Bayesian hierarchical model to normalize the scores across reviewers. Bayesian hierarchical modeling (BHM) is well suited to our small sample scenario because it employs 'partial pooling', a technique in which the model makes more informed estimates for each expert and each question while simultaneously learning from overall patterns across the entire dataset. Essentially, information gleaned from one expert's rating behavior can help refine the understanding of others, and similarly, observed response patterns for some questions can inform estimates related to different questions. This produces stable and uncertainty-aware estimates while adjusting for each rating tendency. No dataset-query combination was repeated to ensure a consistent and unbiased assessment across the four evaluation dimensions.

\subsection{Pilot Session: Exploratory User Interaction}
To gain initial qualitative insight into the system's utility, we conducted a pilot think-aloud session with a robotics research professor unfamiliar with the system's registered datasets. The goal was to observe the participants' exploration strategy,  understand natural interaction patterns, ease of navigating from general to specific information, and the system’s overall effectiveness in finding and revealing the dataset attributes, thereby highlighting both strengths and areas for refinement. After a brief orientation on the system's purpose, we asked the participant to freely interact with the chatbot to find a dataset relevant to his research interest.  This involved initiating broad queries to progressively refining his inquiries to delve deeper into specific characteristics such as provenance, methodology, technical details, and data types.  The session lasted twenty minutes. Observations of interaction sequences, verbalized thoughts, and feedback from a brief post-session discussion were recorded.

\section{Results Summary}
\subsection{Results from the Experts Review}
\begin{figure}[t]
  \centering
  \includegraphics[width=1.0\columnwidth]{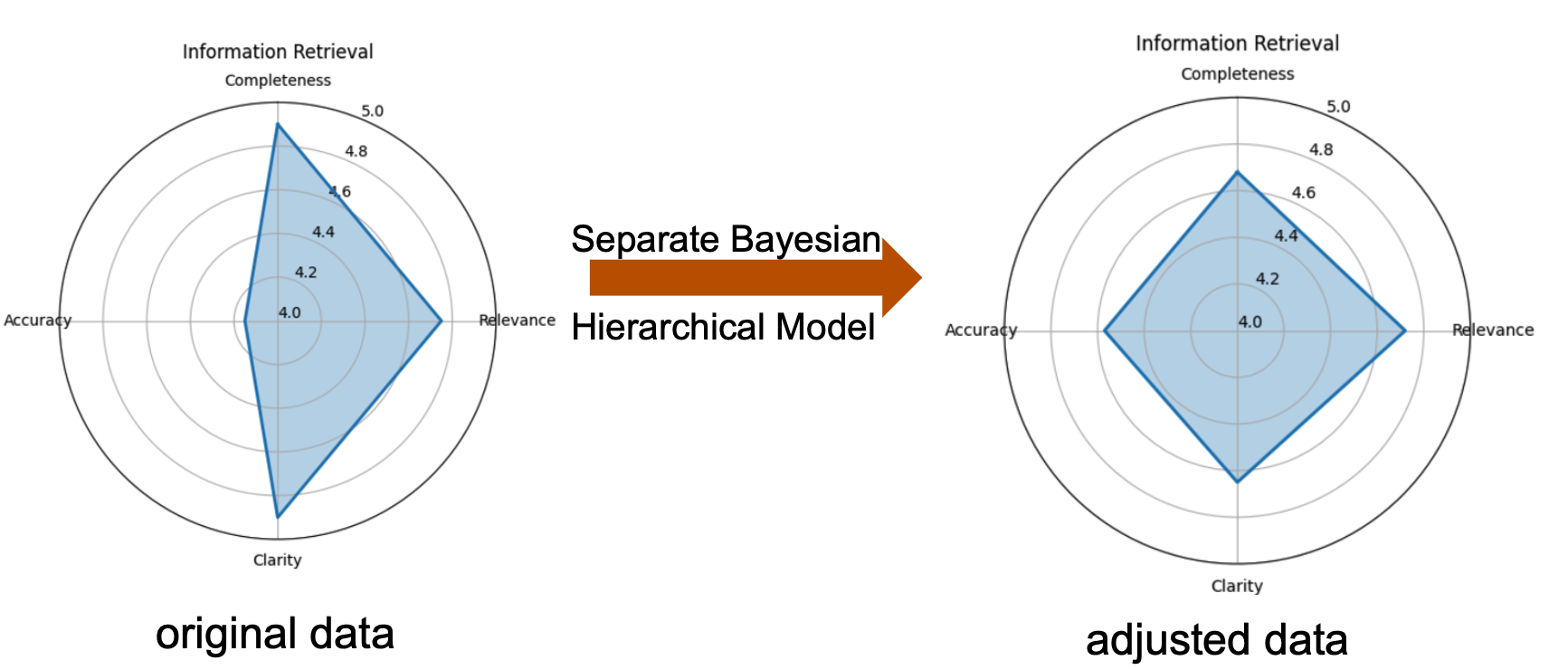}
  \caption{Bayesian Correction in Information Retrieval}
  \label{fig:bayesian_ir}
\end{figure}

The following expert review results detail the chatbot's performance across the aforementioned dimensions. The scores, adjusted for objectivity using the Bayesian Hierarchical Model (BHM), built upon already favorable unadjusted figures, remained strong, consistently reflecting the chatbot's capabilities.
\textbf{1) Information Retrieval:}
The chatbot achieved an average expert rating of 4.65 out of 5, demonstrating consistency in providing structured and relevant answers. Reviewers noted that responses directly referenced precise metadata elements, including links to supporting documentation, questionnaire materials, and descriptions of experimental components. Since the experts preferences mostly varied in this dimension,  we use it to illustrate how BHM corrects for individual bias, as shown in Figure~\ref{fig:bayesian_ir}.

To make the rating process explicit, each expert–prompt score \(y_{ij}\) was modeled as
\[
y_{ij} \sim \mathcal{N}\!\bigl(\mu + \alpha_i + \theta_j + \gamma_{\text{comp}},\, \sigma^{2}\bigr),
\]
Here \(y_{ij}\) denotes the score assigned by expert \(i\) to evaluation prompt \(j\),
with \(i\in\{1,2\}\) (our two raters) and \(j\in\{1,\dots,10\}\) (the ten prompts).

Fitting the model resulted in
\[
\gamma_{\text{comp}} = +0.01
\quad (\text{95\% Credible Interval }[-0.015,\;0.035]).
\]
This indicates that the chatbot's answers were, on average, marginally more complete than the global baseline, confirming that structured metadata enhances the model's precision by enabling it to include the key details researchers expect.

\textbf{2) Answer Stability:}
In this dimension, the chatbot received an average score of 4.9. Responses remained consistent despite variations in query phrasing. Experts highlighted the system's robustness to linguistic variations, which significantly enhances its usability in interdisciplinary research contexts.

\textbf{3) Factual Accuracy:}
The chatbot scored an average of 4.9 in Factual Accuracy. Human experts verified that responses accurately reflected the datasets' documentation, IRB status, robot types, and specific sensor modalities. The evaluation confirmed that the system consistently retrieved accurate, grounded information from the structured metadata and supplementary files.

\textbf{4) Comparison Capability:}
The Comparison capability averaged a score of 4.9. The chatbot effectively identified key differences between the datasets, such as robot control methods and sensory configurations. However, it relies on precise queries that include specific dataset names. General or vague questions (e.g., ``What is the robot model difference?'') often yield poor results, while more targeted queries (e.g., ``What is the robot model difference between CODa and SCAND?'') are handled well. This highlights a key limitation: the system's ability to compare is tied to how well the users can specify their intent. Still, the chatbot successfully utilizes standardized metadata to support structured comparisons, which underscores the utility of the underlying knowledge graph. Recognizing these limitations is crucial for understanding the system's practical use and for guiding future improvements.


Across all four dimensions (Information Retrieval, Answer Stability, Factual Accuracy, and Comparison Capability), the evaluation revealed that the chatbot's strong performance is largely due to the structured metadata foundation. Unlike traditional systems, this meticulously curated, graph-structured knowledge allows the chatbot to interpret precise natural language inquiries.

\subsection{Findings from the Pilot Exploratory Session}
The exploratory session provided valuable qualitative feedback on the system's utility and user interaction. The participant found the conversational interface to be an intuitive starting point for his search process. Throughout the session, the participant was generally able to direct the conversation towards obtaining specific responses regarding the characteristics he was interested in. His feedback likened the system to an "intelligent dataset library," highlighting its effectiveness in helping him narrow the scope of his search and obtain detailed information (such as specific data types or methodological aspects) pertinent to his research goals. He also noted the system's potential to reduce the time and effort associated with searching for academic datasets in comparison to traditional browse and search functions. These positive observations were accompanied by constructive suggestions. For instance, he commented on the need to enhance the clarity and conciseness of some initial chatbot responses, which were occasionally perceived as slightly vague or overly wordy. He also suggested training the system to find models and libraries relevant to robotics research. These insights from the pilot session are being used to guide further iterations of the chatbot interface and interaction design. 

\section{Conclusion and Future Work}
We introduced a prototype system for the FAIR curation, publication, and natural language access of human-centered robotics datasets. The evaluation of our system demonstrated its effectiveness in enhancing HRI datasets' findability, accessibility, interoperability, and reuse. Specifically, findability is ensured through their publication in an institutional repository with persistent digital identifiers; accessibility is improved by combining repository access with scalable online storage; interoperability is supported by a shared data model structured into a knowledge graph; and reuse is achieved through rich metadata, detailed data reports, and clear documentation. These efforts directly align with FAIR principles and address long-standing challenges in robotics data sharing.

The system comprises tightly coupled components. A robust data model, curation best practices, a Neo4j-based knowledge graph, and a sustainable infrastructure including TDR and TACC's cloud and storage resources as the backbone for permanent dataset access. These components work in concert to ensure the long-term availability and usability of ethical HRI data.  In turn, by enabling interactive natural language inquiry and data retrieval, the chatbot highlights and synthesizes the FAIR capabilities and the reliable backend infrastructure. While the chatbot is a powerful tool for sophisticated exploration,  the underlying curation process and infrastructure form the foundation for trustworthy and reusable datasets.

Future work will explore the system’s applicability to broader robotics datasets beyond HRI, and we plan to conduct comparative studies using chatbots trained on datasets of varied curation quality. We also aim to register datasets from multiple repositories. Recognizing the challenge of automating integration across disparate repositories, we will pursue scalable strategies for metadata standardization and ingestion. We will also expand the think-aloud session into a larger, structured study to refine the interaction design. 



\balance

\end{document}